\documentclass[12pt,fleqn]{article}
\input{epsf}
\usepackage{amsmath}
\usepackage{latexsym}
\usepackage{amssymb}

\newlength{\dinwidth}
\newlength{\dinmargin}
\def\simle{\ \lower -2.5pt\hbox{$<$} \hskip-8pt \lower 2.5pt \hbox{$\sim$}\ }
\def\simge{\ \lower -2.5pt\hbox{$>$} \hskip-8pt \lower 2.5pt \hbox{$\sim$}\ }

\begin{document}
\newcommand{\diff}{{\rm d}}
\newcommand{\diffa}{{\rm d}^{3}}
\newcommand{\diffb}{{\rm d}^{4}}
\newcommand{\diffn}{{\rm d}^{n}}
\newcommand{\wt}{Ward-Takahashi identity}
\newcommand{\eu}{{\rm e}}
\newcommand{\im}{{\rm Im}}
\newcommand{\re}{{\rm Re}}
\newcommand{\tr}{{\rm tr}}
\newcommand{\sh}{\not\!}
\newcommand{\tp}{{\rm T}}
\newcommand{\vp}{{\rm P}}
\newcommand{\di}{{\rm Disc}}
\newcommand{\td}{triangular diagram}
\newcommand{\ab}{absorbitive part of the triangular diagram}
\newcommand{\dr}{dispersion relations}
\thispagestyle{empty}
\vspace*{-3cm}
\begin{flushright}
UPRF-99-21\\
December 1999
\end{flushright}
\vspace*{2.5 cm}

\begin{center}
{\huge Axial Anomaly and Polarized Radiative Decays }
\vskip 1.0cm
\begin{large}
{Luca~Trentadue and Michela~Verbeni}\\
\end{large}
\vskip .2cm
 {\it Dipartimento di Fisica, Universit\`a di Parma,\\
 INFN Gruppo Collegato di Parma, 43100 Parma, Italy}\\
\end{center}
\vspace*{.5cm}
 \begin{abstract}
We extend the approach of Dolgov and Zakharov to the axial anomaly to study polarized radiative decays. We analyse the pattern of mass singularity cancellation in the corresponding decay rates. We compare polarized and unpolarized cases. The cancellation of the infrared and collinear singularities is verified to all powers of  the lepton mass for the \(\pi^{+}\) and \(Z^0\) polarized radiative decays. 
\end{abstract}
\vspace*{1cm}
\begin{center}
\end{center}

\newpage
\setcounter{page}{1}

\section{Introduction}\label{Intro}
Long time ago, Dolgov and Zakharov \cite{DZ} have introduced an alternative approach to the axial anomaly \cite{Adler}, based on the study of the triangular diagram, with two vector and one axial vertices, by means of dispersion relations. From this approach, the interpretation of the axial anomaly as an infrared phenomenon follows. 
The infrared aspect of the axial anomaly, arised in this paper, can be seen as complementary to the more familiar ultraviolet one, which emerges from the renormalization procedure. It allows to shed light upon the physical meaning of the chiral symmetry breaking as related to a non conservation of helicity.\\
In \cite{DZ} the absorbitive part of the triangular diagram is computed by making an unitary cut and results proportional to the product of two amplitudes. The first one is relative to the production of a fermion-antifermion pair by an axial source and the second one corresponds to the subsequent annihilation of the pair into two real photons. In both these processes there occur helicity flips, thus the chirality is not conserved in the zero mass limit. For what follows, it is important to examine in particular the second process contributing to the \ab~(\cite{DZ,Huang}).\\
By taking the \(m \to 0\) limit of the \ab, Dolgov and Zakharov find a finite result \cite{DZ}:
\begin{equation}
\im g_1(q^2)\,\longrightarrow \, - \pi \delta(q^2)\;\;\;\;\;\;\;{\rm as}\;m \to 0,
\label{limit}
\end{equation}  
with \(g_1(q^2)\) an invariant scalar function and \(q^2\) the momentum transfer. Eq. (\ref{limit}) shows that the \(m \to 0\) limit is not smooth \cite{Lee}.\\
This behaviour is connected with a non conservation of helicity, in the massless limit, a non conservation of chirality. Thus the axial anomaly emerges by studying the properties of the amplitude in the infrared region. The purpose of this work is to analyse the cancellation of mass singularities in polarized amplitudes. We will not discuss the physical implications of the zero fermion mass limit.\\
As stated in refs. \cite{DZ,Z,S}, the finite result is a consequence of the singularity occurring in the fermion propagator as \(m \to 0\), that exactly cancels the suppression factor \(m^2\) of the product of the amplitudes.\\
The connection between the anomaly and the non conservation of physical quantities has riceved a lot of attention in the literature. 
In a seminal work \cite{Gribov}, Gribov has described the source of the anomalies as a collective motion of particles with arbitrarily large momenta in the vacuum. Related to this work is ref. \cite{Mueller}. The Dolgov and Zakharov approach has been also considered in refs. \cite{S,Falk}.\\
We attempt here to interpret some simple processes, as the \(\pi^{+}\) and \(Z^0\) polarized radiative decays, as a manifestation of the infrared nature of the axial anomaly \cite{carlitz}. In particular, we calculate the relative decay rates, for the cases in which the outgoing leptons are in a definite helicity state and we examine in some detail the structure of the mass singularities and their cancellation. We study how the Kinoshita-Lee-Nauenberg (KLN) theorem \cite{Lee,Kinoshita} applies to these cases and we consider the analogies and the differences with respect to the corresponding unpolarized decay rates.\\

\section{Polarized decay of the pion}\label{Pion}
In the second process contributing to the \ab~a charged fermion with fixed helicity undergoes an helicity flip by emitting a photon \cite{DZ,Huang}. This is a forbidden process in the massless limit, but in this limit the absorbitive part doesn't vanish.\\
We consider helicity changing reactions, studied for the first time by Lee and Nauenberg \cite{Lee}, by using the approach of Dolgov and Zakharov.\\
We have found in the literature other works dealing with this kind of processes. For example in ref. \cite{Falk} the authors consider the process
\begin{equation}
e^{-}(p_{-},\lambda) + A(p)\,\longrightarrow\, e^{-}(p'_{-
},\lambda') +
\gamma(k,\lambda_{\gamma}) + B(q_i), \label{procr}
\end{equation}
where \(A\) is a target, \(\gamma\) is a bremsstrahlung photon, assumed almost collinear with respect to the direction of the incident electron and \(B\) is a set of particles produced in the reaction. \\
For \(\lambda \neq \lambda'\), the electron makes an helicity flip. The helicity flip cross section, computed to the leading order in the electron mass, is given by: 
\begin{equation}
\frac{\diff \sigma_{hf}}{\diff x} = \sigma_0 \left(s(1-x) 
\right)
                     \frac{\alpha}{2 \pi} x, 
 \label{hf}
\end{equation}
where \(s=(p_{-}+ p'_{-})^2\), \(x = k_0/E\), \(E\) is the energy of the incoming electron and \(\sigma_0\) is the cross section for the Born process
\begin{equation}
e^{-}(p_{-} - k,\lambda) + A(p)\,\longrightarrow\,e^{-}(p'_{-
},\lambda) +
B(q_i). \label{procnr}
\end{equation}
We see that the expression (\ref{hf}) does not vanish in the massless limit \cite{Einhorn}.  \\
We now consider the leptonic decay of the pion. At the Born level, the total decay rate relative to the non radiative pion decay
\begin{displaymath}
\pi^{+} \, \longrightarrow\, l^{+} + \nu_l, 
\end{displaymath}
where \({l^{+}}\) is an antilepton (\({e^{+}}\) or \({\mu^{+}}\)) and \({\nu_l}\) is the associated neutrino, is given by:
\begin{equation}
\Gamma_0(\pi^{+}\rightarrow l^{+} \nu_{l}) = 
\frac{G^2 f^{2}_{\pi}}{8 \pi}\, \mid V_{ud}\mid^2 \, \frac{m^2_l}{m^3_{\pi}}\,
(m^2_{\pi} - m^2_l)^2,\label{born}
\end{equation}
where \(G\) is the Fermi coupling constant, \(f_{\pi}\) is the pion decay constant, \(V_{ud}\) is the CKM matrix element, \(m_l\) and \(m_{\pi}\) are the lepton and pion masses, respectively.\\
\(\Gamma_0\) is proportional to \(m^2_l\); this factor is due to the fact, that, due to angular momentum conservation, the pion produces a left-handed lepton, while the structure of the weak coupling requires the \(l^{+}\) to be right-handed for \(m_l=0\).\\
\begin{figure}[t]
\begin{center}
\begin{tabular}{c}
\epsffile{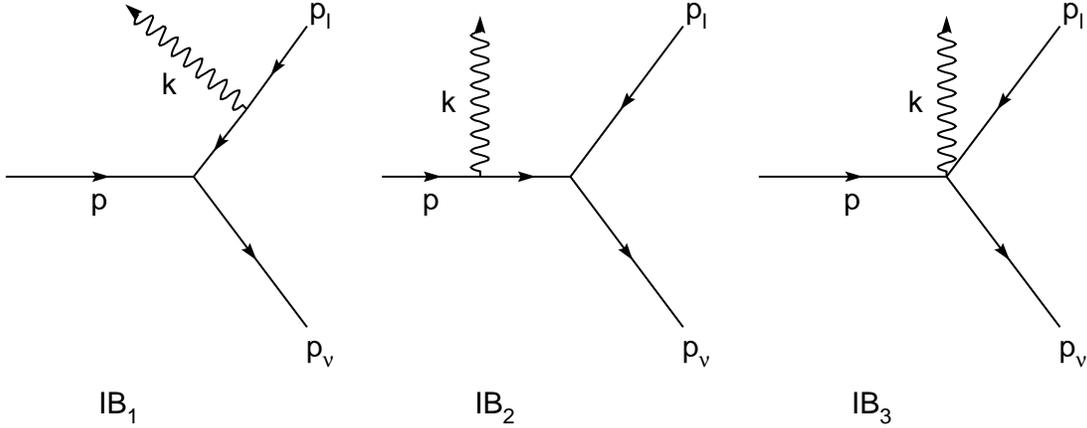}
\end{tabular}
\end{center}
\label{innerb}
\caption{Inner Bremsstrahlung diagrams}
\end{figure}
The total decay rate for the
process in which the lepton is polarized can be written as:
\begin{eqnarray}
\Gamma^{pol}_{0}(\pi^{+}\rightarrow l^{+} \nu_l) = 
\frac{G^2 f^{2}_{\pi}}{16 \pi} \, \mid V_{ud} \mid^2 \, \frac{m^2_l}{m^3_\pi}\,
                     (m^{2}_{\pi} - m^{2}_{l})^2 \,
                     \left(1 - \frac{{\bf p}_l \cdot {\bf s}_l}
                    {|{\bf p}_{l}|} \right),
                       \label{bornpol}
\end{eqnarray}
here \({\bf p}_{l}\) and \(\bf s_l\) are the linear momentum and the spin vector of the lepton, respectively, giving \({\bf p}_l \cdot {\bf s}_l/|{\bf p}_{l}| = \pm 1 \), for right-handed and left-handed lepton, respectively. Eq. (\ref{bornpol}) indicates 
that the lepton is mandatory left-handed, as requested by angular momentum conservation.\\ 
Let us consider the radiative dacay
\begin{displaymath}
\pi^{+} \rightarrow l^{+} + \nu_{l} + \gamma.
\end{displaymath}
The relative amplitude can be written as the sum of two contributions \cite{Bryman}
\begin{equation}
M(\pi^{+}\rightarrow l^{+} \nu_l \gamma) = M_{IB} + M_{SD}, \label{e:4.4}
\end{equation}
where \(M_{IB}\) is the Inner Bremsstrahlung amplitude and \(M_{SD}\) is the Structure Dependent one.\\
We calculate the probability that  the lepton flips its helicity and becomes right-handed, by emitting a real photon. The relevant part, for the problem we are considering, is the Inner 
Bremsstrahlung contribution described by the diagrams of fig. 1.
%\begin{figure}[t]
%\begin{center}
%\begin{tabular}{c}
%\epsfxsize=14truecm
%\epsffile{innerb.eps}
%\end{tabular}
%\end{center}
%\label{innerb}
%\caption{Inner Bremsstrahlung diagrams}
%\end{figure}                   
The \(IB_3\) diagram gives the so called contact term, introduced to ensure gauge invariance (see for example \cite{Bryman}). For the moment, we consider only tree diagrams. At the first order in perturbation theory, we retain terms of all powers in the lepton mass. We will argue that the manifestation of the axial
anomaly is strictly connected to these terms.\\
The Inner Bremsstrahlung amplitude is \cite{Bryman}:
\begin{eqnarray}
M_{IB} & = & \sum_{i=1}^{3}\, IB_{i}  \nonumber \\
       & = & i e \, \frac{G}{\sqrt{2}}\, f_{\pi} \, V_{ud}\, m_l\,
              \bar{u}(p_{\nu}) (1 + \gamma_5) 
          \left[\frac{p \cdot \epsilon}{p \cdot k} - 
                \frac{\not\!k \not\!\epsilon + 2 p_l \cdot \epsilon}
                     {2 p_l \cdot k} \right] v(p_l, s_{lR}), \label{e:4.5}
\end{eqnarray}
where \(u\) and \(v\) are the Dirac spinors for the neutrino and the lepton, \(p\) is the pion momentum, \(p_l\) and \(s_{lR}\) are the momentum and the polarization vector of the right-handed lepton, \(k\) and 
\(\epsilon\) are the momentum and the polarization vector of the 
photon, respectively.\\
 We see that \(M_{IB}\) is proportional to the lepton mass \(m_l\), thus the 
decay rate is proportional to \(m^2_l\). As we have said above, this factor is a consequence of the structure of the weak current. We remove it by normalizing the radiative decay rate with respect to the 
non radiative one.\\
The differential Inner Bremsstrahlung 
contribution for right-handed lepton is:
\begin{eqnarray}
& &\frac{1}{\Gamma_0}\,\frac{\diff \Gamma^{R}_{IB}}{\diff y} =
\frac{\alpha}{4 \pi}\,\frac{1}{(1-r)^2} \, \frac{1}{A(1-y+r)}
\bigg\{2A \, \Big[1+y^2 - 2A + r(2A+r-6)\Big] \nonumber \\
& & + \,\Big[(A+2r)(1+r)^2 + Ay(y-4r) + 2ry(1+y)-y(1+y^2+5r^2)\Big]\,\ln{\frac{y+A}{y-A}} \nonumber \\
& & + \,(1-y+r)^2(y-2r-A) \ln{\frac{y+A-2}{y-A-2}}\bigg\}.
\label{IBRa}
\end{eqnarray}
The dimensionless variable \(y\) is defined as 
\begin{displaymath}
y \, =\, \frac{2E_l}{m_{\pi}}
\end{displaymath}
and 
\begin{displaymath}
r \, =\, \frac{m^2_l}{m^2_{\pi}},\;\;\;\;\;\;A \, =\,\sqrt{y^2-4r}.
\end{displaymath}
The physical region for \(y\) is:
\begin{equation}
2\sqrt{r} \le y \le 1+r.
\label{region}
\end{equation}
We see that, in eq. (\ref{IBRa}), there is a term independent of the lepton mass, the one in the first square brackets. Owing to this term, the differential decay rate 
doesn't vanish in the limit \(m_l \rightarrow 0\). We have:
\begin{equation}
\frac{1}{\Gamma_0} \frac{\diff \Gamma^R_{IB}}{\diff y}\,
\longrightarrow \, \frac{\alpha}{2 \pi} \, (1-y)\;\;\;\;\;\;\;\;{\rm as}\;r \to 0.\label{massa0}
\end{equation}
This indicates that there occurs an helicity flip and thus a chirality non conservation in the limit of zero lepton mass. According to the interpretation given above, 
this term can be interpreted as connected to the axial anomaly. It corresponds to the anomalous term present in the divergence of the axial current.\\
Since the radiative process with the right-handed \(l^{+}\) is forbidden in the limit
 \(m_l \rightarrow 0\) by the chiral invariance of the massless QED
Lagrangian, the result different from zero, in this limit, indicates the occurrence of a cancellation mechanism, analogous to the one acting in the absorbitive part of the triangular diagram. Indeed, to obtain the decay rate in eq. (\ref{IBRa}), we have integrated over the emission angle terms containing the lepton propagator squared; this produces power collinear singularities, exactly cancelling the chiral suppression factor \(m^2_l\) and thus producing the term related to the axial anomaly. The integration over the lepton propagator produces logarithmic collinear singularities.\\

\section{Mass singularities}\label{Mass}
In this section we discuss the structure of mass singularities and their cancellation in the decay rate for the radiative pion and \(Z^0\) decays and how the KLN theorem applies to these cases.\\

\subsection{The pion case}\label{Pion case}
It is useful to separate the cases of unpolarized, right-handed and left-handed outgoing lepton. \\
Let us consider the mass singularity cancellation for the familiar case of unpolarized radiative \(\pi^{+}\) decay to the first order in \(\alpha\).
The expression for the Inner Bremsstrahlung contribution, differential with 
respect to the lepton energy, is given by:
\begin{eqnarray}
\! \frac{1}{\Gamma_0}\,\frac{\diff \Gamma_{IB}}{\diff y} & = & 
\frac{\alpha}{4 \pi}\,\frac{1}{(1-r)^2} \, \frac{1}{(1-y+r)}\,
\bigg\{4A(r-1) + \Big[(1+r)^2 + y(y-4r)\Big]\,\ln{\frac{y+A}{y-A}} \nonumber \\
& - & (1-y+r)^2\,\ln{\frac{y+A-2}{y-A-2}} \bigg\}.
\label{IRnpa}
\end{eqnarray}
One can easily see that the eq. (\ref{IRnpa}) is divergent both in the collinear and in the infrared limits. The coefficients of the collinear logarithms don't go to zero in the limit \(r \to 0\). There are also infrared divergences, since if we let \(y\) reach its kinematical limit \(y^{MAX} = 1+r\),
corresponding to the photon energy going to zero, the expression (\ref{IRnpa}) diverges.\\
The decay rate is made free from mass singularities in the ordinary way: the divergences cancellation occurs 
in the total inclusive decay rate, when we add all the first order contributions to the 
perturbative expansion, i.e. those relative to real and virtual photon emission.\\
The diagrams describing the real photon emission were already given in fig. 1; the 
diagram for the virtual correction is drawn in fig. 2.
\begin{figure}[t]
\begin{center}
\begin{tabular}{c}
\epsffile{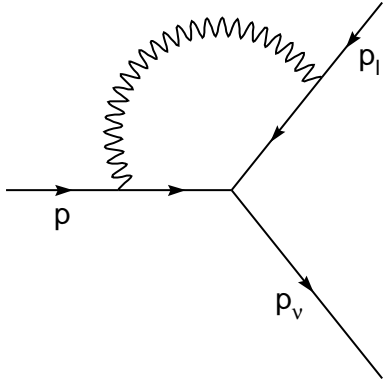}
\end{tabular}
\end{center}
\label{fig5}
\caption{Virtual diagram}
\end{figure}
The expression for the lepton energy spectrum, including the Inner Bremsstrahlung contribution and the virtual photon one, calculated to the leading order in  \(m_l/m_{\pi}\), is given by \cite{Kuraev}:
\begin{equation}
\frac{1}{\Gamma_0} \frac{\diff \Gamma}{\diff y} = 
D(y,r)\left[1 + \frac{\alpha}{\pi} K_l(y)\right].  \label{spettroy}
\end{equation} 
\(D(y,r)\) is the lepton distribution function, given, to the first order in \(\alpha\), by:
\begin{equation}
D(y,r) = \delta (1-y) + \left[ \frac{\alpha}{\pi}(L-1) + O(\alpha^2)\right]P^{(1)}(y),
\label{defD}
\end{equation}
where \(L\) is the logarithm
\begin{equation}
L = \ln{\frac{m_{\pi}}{m_l}},
\label{defL}
\end{equation}
diverging in the collinear limit; if the lepton is an electron, \(L \simeq 5.6\).
\(P^{(1)}(y)\) is the Gribov-Lipatov-Altarelli-Parisi kernel \cite{Altarelli}, which can be expressed in the form:
\begin{equation}
 P^{(1)}(y) = \frac{1+y^2}{1-y} - \delta (1-y) \int^1_0 \diff z 
\frac{1+z^2}{1-z}.
\label{defP}
\end{equation}
\(K_l(y)\) is a finite term, free from infrared and collinear singularities, which has the expression:
\begin{equation}
K_l(y) = 1 - y - \frac{1}{2}(1-y)\ln{(1-y)} +
\frac{1+y^2}{1-y}\ln{y}.
\label{defK}
\end{equation} 
 The differential decay rate to order \(\alpha\) therefore becomes:
\begin{equation}
\frac{1}{\Gamma_0}\, \frac{\diff \Gamma}{\diff y} = 
\delta(1-y) + \frac{\alpha}{\pi}\,(L-1)\,P^{(1)}(y) + 
\frac{\alpha}{\pi}\,K_l(y).
\label{ordalfa}
\end{equation}
To calculate the inclusive decay rate, we have to integrate the expression in eq. (\ref{ordalfa}) over \(y\); to the leading order in the lepton mass, the physical region for y is \(0 \le y \le 1\). The kernel \(P^{(n)}\) has the property that:
\begin{equation}
\int^1_0 \diff y P^{(n)}(y) = 0;
\label{canc}
\end{equation}
thus, when we calculate the inclusive decay rate, the coefficient of the collinear logarithm vanishes and the resulting expression is finite in the zero mass limit.\\
Carring out the integration over \(y\), we obtain the well known inclusive decay rate to order \(\alpha\)
\begin{equation}
\frac{\Gamma}{\Gamma_0} = 1 + 
\frac{\alpha}{\pi}\,[\frac{15}{8} - \frac{\pi^2}{3}].
\label{tot}
\end{equation}
As expected, the expression (\ref{tot}) is finite in the collinear limit and is also free from infrared divergences, because, as usual, the infrared divergences present in the soft photon contribution and in the virtual photon contribution have cancelled each other.\\
Let us now discuss the mass singularities in the case of the right-handed Inner Bremsstrahlung contribution, given in eq. ({\ref{IBRa}}). We have already seen in section 2 that \(\diff \Gamma^R_{IB}/\diff y\) is finite in the limit \(r \to 0\), 
i.e. it is free from collinear singularities. In this limit the coefficients 
of both the collinear logarithms vanish. Indeed, as we have discussed above, 
only the term related to the axial anomaly survives in the zero mass limit, that is the part of the decay rate indipendent of 
the lepton mass.\\
We observe that the right-handed Inner Bremsstrahlung contribution is free from infrared divergences as well. If we make the lepton energy 
\(y\) reach its kinematical limit \(y^{MAX}\), we obtain a finite result:
\begin{eqnarray}
\frac{1}{\Gamma_0}\frac{\diff \Gamma^R_{IB}}{\diff y} \longrightarrow  0\;\;\;\;\;\;\;{\rm as}\;y \to y^{MAX}.
\label{limIR}
\end{eqnarray}
The result ({\ref{limIR}}) shows that the emission of soft photons does not contribute to the radiative \(\pi^{+}\) decay with a right-handed lepton. Indeed, the soft photon contribution factorizes with respect to the Born decay rate, but this vanishes in the case of right-handed \(l^{+}\) (see eq. (\ref{bornpol})). Physically, eq.(\ref{limIR}) is a consequence of the fact that soft photons don't carry spin, thus they cannot contribute to the angular momentum balance; therefore the process with the right-handed lepton emitting a soft photon, is forbidden by angular momentum conservation.\\
For the same reason of angular momentum conservation, in the right-handed case also the virtual contribution is zero.
The virtual correction is factorized with respect to the Born decay rate, but, as we have already seen, if the lepton is right-handed, this is identically 
zero.\\
In the right-handed case, the mass singularity cancellation occurs through a mechanism different from the one working in the unpolarized decay rate. The infrared and the collinear limits give, separately, a finite result. In particular, the coefficient of the collinear logarithms is the lepton mass, instead of the usual correction factor coming from the soft and the virtual photon contributions, as in eq. (\ref{ordalfa}).\\
This particular mass cancellation mechanism is the consequence of the combination of two constraints: the angular momentum conservation in the pion vertex and the helicity flip in the photon-lepton vertex.\\
The situation is completely different if we consider the radiative process with the outgoing left-handed lepton, i.e. the process without helicity flip. The left-handed Inner Bremsstrahlung contribution is given by:
\begin{eqnarray}
& &\frac{1}{\Gamma_0}\frac{\diff \Gamma^L_{IB}}{\diff y} = 
\frac{\alpha}{4\pi} \, \frac{1}{(1-r)^2} \, \frac{1}{A(1-y+r)}
\bigg\{2A\, \big[r(2A-r+6) -y^2 -1-2A\big] \nonumber \\
& & + \big[(A-2r)(1+r)^2 + Ay(y-4r) -2ry(1+y) + y(5r^2+y^2+1)\big]
\, \ln{\frac{y+A}{y-A}} \nonumber \\
& & + (1-y+r)^2(2r-A-y) \, \ln{\frac{y+A-2}{y-A-2}} \bigg\}
\label{IBLa}
\end{eqnarray}
The expression (\ref{IBLa}) contains collinear singularities, since the coefficients of the collinear logarithms don't vanish in the limit \(r \to 0\). It is also infrared divergent, as one can verify by taking the limit \(y \to y^{MAX}\). In this case we don't have the constraint constituted by the helicity flip in the photon-lepton vertex and in the pion vertex the angular momentum is conserved for soft and virtual photon emission. Thus, the mass singularity cancellation mechanism 
occurs in the ordinary way, as in the unpolarized case, i.e. in the total inclusive decay rate, obtained by adding all the order $\alpha$ contributions. \\
Let us show how the cancellation takes place. As we have already seen (eq. (\ref{bornpol})), in the Born \(\pi^{+}\) decay the outgoing lepton is left-handed, due to angular momentum conservation. 
Thus the Born decay rates with unpolarized and left-handed \(l^{+}\) coincide:
\begin{equation}
\Gamma_0^L = \Gamma_0.
\label{0L0np}
\end{equation}
Because of the factorization with respect to the Born decay rate, also the unpolarized and the left-handed virtual contributions are equal:
\begin{equation}
\Gamma^L_v = \Gamma_v. 
\label{vLvnp}
\end{equation}
Expressing the left-handed Inner Bremsstrahlung contribution in terms of the unpolarized and the right-handed ones, the total contribution to order \(\alpha\) to the left-handed process is given by:
\begin{equation}
\Gamma^L_{TOT} = (\Gamma_0 + \Gamma_v + \Gamma_{IB})
               - \Gamma^R_{IB}.
\label{TOTL}
\end{equation}
The expression (\ref{TOTL}) is finite both in the infrared and in the collinear limit, because the mass singularities present in the terms between brackets cancel each other, as we have seen (eq. (\ref{tot})) and \(\Gamma^R_{IB}\) is free from mass singularities.\\
The presence of mass singularities is a conseguence of the fact that the states of a theory containing massless particles are highly degenerate.
The KLN theorem states that the mass singularities disappear from the transition probability when we average it over the ensemble of degenerate states. 
It results that imposing on the outgoing lepton a polarization opposed to the one prescribed by the interaction taking place before the photon emission, implies a reduction of the degeneration subspace \cite{Trentadue}. Thus in this case we have a particular application of the KLN theorem.  \\

\subsection{The \(Z^0\) case}\label{Z}
Let us start the discussion on the \(Z^0\) radiative decay by saying that it differs from the pion case. 
For radiative pion decay, due to the angular momentum conservation in the pion vertex, there is no room for a right-handed lepton. For such a channel, soft and virtual photon contribution give a contribution equal to zero. This result is valid independently of the lepton mass.
Let us now consider a more general process, by loosing the value of the angular momentum of the decaying state. As an example, we study the radiative \(Z^0\) decay in a lepton-antilepton \((l^{-}l^{+})\) pair, in which the lepton is in a definite helicity state.\\
The \(Z^0\)-leptons vertex is:
\begin{displaymath}
i \, \frac{M_Z}{\sqrt{2}}\,\left(\frac{G}{\sqrt{2}}\right)^{1/2}\, 
\gamma_{\mu}(g_v - g_a \gamma_5),
\end{displaymath}
with
\begin{displaymath}
g_v = 1-4 \sin{\theta_W}^2 \;\;\;\;\;\;\;g_a= 1
\end{displaymath}
where \(\theta_W\) is the Weinberg angle and \(M_Z\) is the \(Z^0\) mass.\\
If we set \(g_v=g_a=1\), we require that in the limit of zero lepton mass, the \(Z^0\) couples to a left-handed lepton. We calculate the decay rate for the process in which the outgoing lepton is right-handed.\\
 At the Born level this is given by:
\begin{eqnarray}
\Gamma^R_0 & = & \frac{GM^3_Z}{48\sqrt{2}\pi}\, \bigg\{\sqrt{1-4r}\,
\big[(g^2_v+g^2_a)(1-r) + 3r(g^2_v-g^2_a)\big]
- 2g_vg_a(1-4r) \bigg\}.
\label{bornz}
\end{eqnarray}
If in eq. (\ref{bornz}) we set \(g_v=g_a=1\), \(\Gamma^R_0\) vanishes in the chiral limit, since there isn't the term related to the axial anomaly.\\
Let us now study the decay process with the lepton emitting a real photon (see fig. 3)
\begin{figure}[t]
\begin{center}
\begin{tabular}{c}
\epsffile{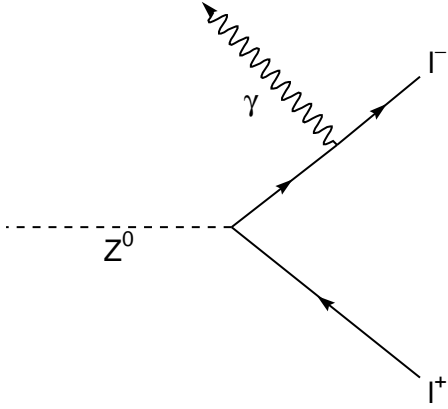}
\end{tabular}
\end{center}
\label{fig6}
\caption{Diagram contributing to the decay rate \ref{Rz}}
\end{figure}
and evaluate the probability that the outgoing lepton is right-handed. The electromagnetic interaction doesn't couple states with different chirality, hence we expect the decay rate to vanish for \(m_l \to 0\).\\
The decay rate for the process described by the diagram of fig. 3, differential with respect to the lepton energy, is given by the expression:
\begin{eqnarray}
& &\frac{\diff \Gamma^R_{\gamma}}{\diff y} = 
\frac{G \alpha M^3_Z}{96 \sqrt{2}\pi^2}\,
\bigg\{\frac{(y-2)(1-y)^2}{4(1-y+r)^2}\,
\Big[(g^2_v+g^2_a)A + 2g_vg_a(2r-y)\Big] \nonumber \\
& + & \frac{(1-y)}{2(1-y+r)}\,
\Big[(g^2_v+g^2_a)A(2r-y) + 2g_vg_a(y^2-2r)\Big] \nonumber \\
& + & \frac{2}{(y-1)}\,
\Big[2(g^2_v-2g^2_a)Ar + (g^2_v+g^2_a)A + 2g_vg_a(4r-y^2+y-1)\Big] \nonumber \\
& + &\Big[g^2_v+g^2_a + 2g_vg_a\,\frac{1}{A}\,\big(4r^2+r(y-y^2+1)-y\big)\Big]
\,\Big[\ln{\frac{y+A}{y-A}} - \ln{\frac{y+A-2}{y-A-2}}\Big]\bigg\}. \nonumber \\
\label{Rz}
\end{eqnarray}
Here
\begin{displaymath}
r=\frac{m^2_l}{M^2_Z}
\end{displaymath}
and \(y\) is the usual dimensionless variable:
\begin{displaymath}
y=\frac{2E_1}{M_Z}
\end{displaymath}
where \(E_1\) is the lepton energy and
\begin{displaymath}
A=\sqrt{y^2-4r}.
\end{displaymath}
The physical region for \(y\) is
\begin{equation}
2\sqrt{r} \le y \le 1.
\label{regionz}
\end{equation}
The result obtained, as given by the emission of the photon by a single leg, is gauge dependent. To have a gauge independent amplitude, the contribution of the diagram b) of fig. 4 must be added \cite{Trentadue}. For the purpose of the polarized amplitude, however, the helicity flip contribution of the diagram b) of fig. 4 gives zero in the massless limit and is therefore negligible in our discussion.\\
From now on we set \(g_v=g_a=1\), to have the condition of chirality conservation in the \(Z^0\) vertex for \(m_l \to 0\). Taking this limit in eq. (\ref{Rz}), we see that \(\diff \Gamma^R_{\gamma}/\diff y\) does not vanish:
\begin{equation}
\frac{\diff \Gamma^R_{\gamma}}{\diff y} \longrightarrow 
\frac{G\alpha M^3_Z}{24\sqrt{2}\pi^2}\,(1-y)\;\;\;\;\;{\rm as}\;r \to 0,\;g_v=g_a=1.
\label{collz}
\end{equation}
The result of this limit is the contribution related to the axial anomaly, which has the same form of the one found in the pion case.\\
Let us now discuss the mass singularity cancellation mechanism for the \(Z^0\) radiative decay case. Eq. (\ref{collz}) shows that the collinear limit gives a finite result. Thus, we conclude that, as in the case of the radiative pion decay with right-handed lepton, the collinear and infrared limits are disconnected.\\
 If we keep the lepton mass different from zero, the chirality is not fixed by the interaction occurring before the photon emission, even if we set \(g_v=g_a=1\). Thus, for \(m_l \neq 0\), the soft and virtual photon contributions are different from zero and diverge in the infrared limit.
If we let the lepton energy reach its kinematical limit, \(y^{MAX}= 1\), we see that \(\diff \Gamma^R_{\gamma}/\diff y\) diverges.
We expect the infrared divergences to cancel, if we add all the first order contributions, given by the diagrams of fig. 3 and 4 
\begin{figure}[t]
\begin{center}
\begin{tabular}{c}
\epsffile{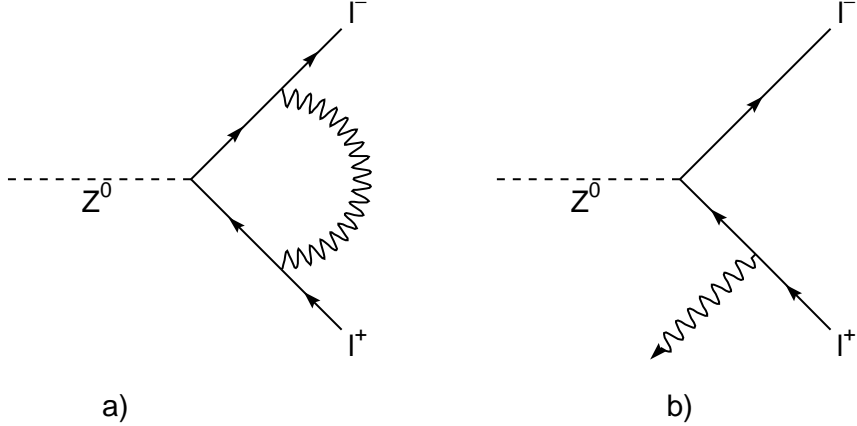}
\end{tabular}
\end{center}
\label{fig7}
\caption{The other order $\alpha$ contributions}
\end{figure}
and calculate the totally inclusive decay rate.\\
If we take the limit \(y \to 1\) in eq. (\ref{collz}), we obtain:
\begin{equation}
\frac{\diff \Gamma^R_{\gamma}}{\diff y} \longrightarrow 0
\;\;\;\;\;{\rm as}\;m_l\to 0, \;y \to 1 \;{\rm and}\;g_v=g_a=1.
\label{IRz}
\end{equation}
Eq. (\ref{IRz}) indicates that, in the massless limit, the soft photon contribution is zero. Indeed, it is factorized with respect to the Born decay rate, which, for \(r \to 0\) and \(g_v=g_a=1\), vanishes. As we have discussed in section \ref{Intro}, the presence of the term related to the axial anomaly is directly connected to the emission of the photon, hence it vanishes in the limit in which the photon is not radiated.\\ 
The virtual photon contribution vanishes in the zero mass limit, as well. Indeed, it is factorized respect to the Born decay rate, which goes to zero as \(r \to 0\). The virtual correction factor can produce only a logarithmic collinear singularity, not a power-like one, needed for the cancellation of the chiral suppression.\\
In the massless limit, also the diagram with the photon emitted by \(l^{+}\) doesn't contribute, since, as the virtual contribution, cannot produces a power type collinear singularity.\\
Taking the infrared limit \(x \to 0\) in eq. (\ref{hf}), gives a finite result (indeed the cross section vanishes). This is a consequence of the fact that the cross section (\ref{hf}) has been calculated to the leading order in the lepton mass. From the eq. (\ref{Rz}), we see that the infrared divergent term is the one given by the term:
\begin{equation}
\frac{4r}{(y-1)} \Big[(g^2_v - 2g^2_a)A + 4g_vg_a \Big] \nonumber
\label{IRd}
\end{equation}
i.e., it is proportional to the lepton mass. Performing the calculation, neglecting the mass terms as done in ref. \cite{Falk}, means imposing the chirality conservation law in the \(Z^0\) vertex; thus the soft photon contribution is zero and the infrared divergences disappear. To the leading order in the lepton mass, one has (as in eq. \ref{hf}) only the term related to the axial anomaly, which vanishes in the infrared limit.\\

\section{Conclusions}\label{Conclus}
We have seen that the decay rates for processes with a fermion changing helicity by emitting a massless vector boson don't vanish in the chiral limit. We have interpreted this behaviour as related to the axial anomaly, as first suggested by Dolgov and Zakharov.\\
In the processes we have considered, the cancellation of the collinear singularities occurs through a mechanism different from the one for the unpolarized case. 
However, we have found a difference between the pion case and the more general case of the \(Z^0\) decay. Due to the angular momentum conservation, the virtual and soft photon contributions are zero, even if the lepton mass is kept different from zero, giving finite infrared and collinear limits. \\
In the case of the radiative \(Z^0\) decay, the decay rate is not finite in the infrared limit, since, for \(m_l \neq 0\), the real soft photon contributions are different from zero. 
We observe that, in order to make the collinear limit finite, it is sufficient to sum over the set of degenerate states containing the fermion making the helicity flip accompanied by a hard collinear photon \cite{Trentadue}.\\
In these processes the collinear limit results disconnected from the infrared one, in the sense that the virtual photon contribution is not needed to render this limit finite.\\
We conclude that the collinear singularity cancellation mechanism is controlled by the anomalous breaking of the chiral symmetry. The axial anomaly implies that taking the collinear limit gives a finite result, independent of the fermion mass.\\
These results can be applied to interpret the mass singularity cancellation mechanism for any polarized process. The extension of this approach to the case of Quantum Cromodynamics for polarized processes is under study.

\newpage
%\vspace{5mm}
\begin{center}
\begin{large}
{\bf Acknowledgments}
\end{large}
\end{center}
\noindent We wish to thank E. Kuraev and V. Fadin for valuable comments and S. Forte, J. Kodaira and L. Lipatov for useful discussions.

\end{document}